\documentclass[conference]{IEEEtran}
\usepackage[american]{babel}
\usepackage[
backend=biber,
style=ieee,
natbib=true,
doi=false,
isbn=false,
eprint=false,
url=true,
uniquelist=false,
mincitenames=1,
maxcitenames=1,
urldate=long
]{biblatex}
\DeclareSourcemap{
    \maps[datatype=bibtex]{
        \map[overwrite] {
            \step[fieldset=abstract, null]
            \step[fieldset=note, null]
            \step[fieldset=pages, null]
        }
    }
}
\AtBeginBibliography{\scriptsize}
\usepackage{amsmath,amssymb,amsfonts}
\usepackage{algorithmic}
\usepackage{graphicx}
\usepackage{textcomp}
\usepackage[usenames,dvipsnames,svgnames,table]{xcolor}
\usepackage{booktabs}
\usepackage[hidelinks]{hyperref}
\usepackage{svg}

\graphicspath{{./}{./images/}}

\IfFileExists{/dev/null}{%
  }{%
}

\usepackage{listings}
\usepackage[noabbrev,capitalise]{cleveref}

\usepackage{microtype}

\usepackage[T1]{fontenc}

\DeclareUnicodeCharacter{2212}{\textendash} 

\usepackage{siunitx}
\usepackage{url}

\usepackage{caption}
\usepackage{subcaption}

\usepackage[a4paper, total={184mm,239mm}]{geometry}
\def\BibTeX{{\rm B\kern-.05em{\sc i\kern-.025em b}\kern-.08em
    T\kern-.1667em\lower.7ex\hbox{E}\kern-.125emX}}

\usepackage{csquotes}
\usepackage{orcidlink}

\svgpath{{images/}}

\usepackage{array, makecell}
\usepackage{fancyhdr}

\begin{document}

\title{Large Language Models to Generate System-Level Test Programs Targeting Non-functional Properties}

\author{\IEEEauthorblockN{Denis Schwachhofer\textsuperscript{1,4}\orcidlink{0000-0002-6763-2948},
Peter Domanski\textsuperscript{2},
Steffen Becker\textsuperscript{1},
Stefan Wagner\textsuperscript{1,5},\\
Matthias Sauer\textsuperscript{3},
Dirk Pflüger\textsuperscript{2},
Ilia Polian\textsuperscript{4}}
\\[0.0cm]
\IEEEauthorblockA{\textsuperscript{1}Institute of Software Engineering, University of Stuttgart, Stuttgart, Germany}
\IEEEauthorblockA{\textsuperscript{2}Institute for Parallel and Distributed Systems, University of Stuttgart, Stuttgart, Germany}
\IEEEauthorblockA{\textsuperscript{3}Advantest Europe, Boeblingen, Germany}
\IEEEauthorblockA{\textsuperscript{4}Institute of Computer Engineering and Computer Architecture, University of Stuttgart, Stuttgart, Germany}
\IEEEauthorblockA{\textsuperscript{5}Technical University of Munich, Heilbronn, Germany}
}

\maketitle
\thispagestyle{fancy}
\fancyhead{}
\fancyfoot{}
\renewcommand{\headrulewidth}{0pt}
\renewcommand{\footrulewidth}{0pt}
\fancyfoot[L]{Testmethoden und Zuverlässigkeit von Schaltungen und Systemen, TuZ 2024.}

\begin{abstract}
System-Level Test (SLT) has been a part of the test flow for integrated circuits for over a decade and still gains importance. 
However, no systematic approaches exist for test program generation, especially targeting non-functional properties of the Device under Test (DUT).
Currently, test engineers manually compose test suites from off-the-shelf software, approximating the end-user environment of the DUT.
This challenging and tedious task does not guarantee sufficient control over non-functional properties.
This paper proposes Large Language Models (LLMs) to generate test programs. 
We take a first glance at how pre-trained LLMs perform in test program generation to optimize non-functional properties of the DUT.
Therefore, we write a prompt to generate C code snippets that maximize the instructions per cycle of a super-scalar, out-of-order architecture in simulation.
Additionally, we apply prompt and hyperparameter optimization to achieve the best possible results without further training.

\end{abstract}


\section{Introduction}\label{sec:introduction}
System-Level Test (SLT) is increasingly used to improve the quality assurance of complex Systems-on-Chip.
In SLT, the Device under Test (DUT) is placed into an environment that emulates its end-user environment as closely as possible.
SLT is executed by running off-the-shelf software in the DUT's mission mode~\cite{Chen2018}.
As such, SLT programs are expected to exercise execution paths and transactions unlikely to be triggered by structural tests, detecting defects that are not covered otherwise~\cite{Polian2020}.
Currently, test engineers manually compose the test suite based on field returns and personal experience with system engineers.
However, this process is challenging and time-consuming, especially given that structural information about the DUT is unavailable.

For SLT, we consider the DUT to be a black box.
This constraint holds since the DUT might contain locked intellectual property, and thus netlists are not available.
Additionally, the complexity of the DUT paired with the length of typical SLT programs makes traditional methods, such as automatic test pattern generation and fault simulation, infeasible.
As such, we cannot use traditional fault models and metrics such as fault coverage to determine the quality of SLT programs.

Non-functional properties, e.g., temperature, play a crucial role in SLT.
The detection of marginal defects depends on the operating conditions of the DUT~\cite{Chen2018}.
For example, there might be a defect on the bus between two modules, which can only be triggered if there is a certain temperature gradient between these two modules.
On the other hand, using functional patterns can help to prevent damage to DUTs due to thermal runaway caused by excessive switching activity.

Instead of directly measuring the temperature, we use another metric as a proxy: Instructions per Cycle (IPC).
IPC tells us how much the pipeline and the functional units are utilized.
A full pipeline consumes more power as there is a higher switching activity~\cite{Hadjilambrou19}, which results in a higher temperature.

This paper proposes using Large Language Models (LLMs) to automatically generate SLT programs that target non-functional properties of the DUT.
The goal is to demonstrate the feasibility of LLMs as code generators in SLT by generating code snippets that maximize the IPC of a super-scalar, out-of-order processor called BOOM~\cite{Zhao2020}.
For this purpose, we examine Code Llama~\cite{roziere2023code}.
We developed a prompt that tells Code Llama to generate C code and gives it some constraints on the output format and contents.
Furthermore, we also use prompt and hyperparameter optimization to achieve the best possible results without fine-tuning the LLM.

\section{Related Work}\label{sec:relatedwork}
In this section, we introduce some related work for functional test generation in the context of hardware testing.
Afterward, we list some related work regarding LLMs as an optimizer and for code synthesis.

\subsection{Test Generation for System-Level Test}\label{subsec:testgeneration}
There is some work by~\citet{DELIGIANNIS_TESTPROGRAMGENERATIONSTRESS_MM_2023, Deligiannis2021a} using formal methods to generate small snippets that maximize the switching activity in specific modules in a processor.
They aim to generate small code snippets for Software-based Self-test using MaxSAT and Genetic Programming.
\citet{Riefert2016} introduce a framework based on an automatic test pattern generator and bounded model checking that is able to generate functional tests to detect small delay defects.

In previous work, we have shown that mutation-based greybox fuzzing can be used to generate functional test patterns~\cite{Schwachhofer2023}.
This work aimed to maximize the power consumption of the BOOM core running on an FPGA.

The aforementioned methods are based on assembly language.
In this work, we instead target a higher-level language, namely C.
Furthermore, to the best of our knowledge, we are the first to use LLMs in the context of hardware testing.

\subsection{Large Language Models as Optimizers and Code Generators}\label{subsec:background}

Recently,~\citet{yang2023large} proposed to use LLMs as iterative optimizers, e.g., in derivative-free optimization problems.
Instead of following previous works for automatic prompt generation~\cite{zhou2022large}, the proposed method constructs the prompt using two core pieces of information: previously generated outputs (including corresponding scores) and the optimization problem description.
The proposed prompt generation scheme allows creating of new prompts in each optimization step and thus iteratively increases the test accuracy based on preceding prompts.
Therefore, prompting and using LLMs similarly could be highly beneficial if we aim to optimize the non-functional properties of a DUT, such as maximizing instructions per cycle.

Moreover, LLMs are frequently used for code generation.
Therefore, LLMs leverage pre-training and fine-tuning approaches to understand and generate human-like code across programming languages.
They have proven successful in various tasks like code completion or code generation from natural language descriptions.
A similar research direction is summarized in a recent survey.
\citet{wang2023software} show how LLMs are used for software test generation, including unit test case generation or test oracle generation.
The methodologies applied effectively in early works provide a valuable starting point for generating SLT programs from natural language descriptions.

\begin{figure*}[!htp]
    \centering
    \includegraphics[width=0.9\textwidth]{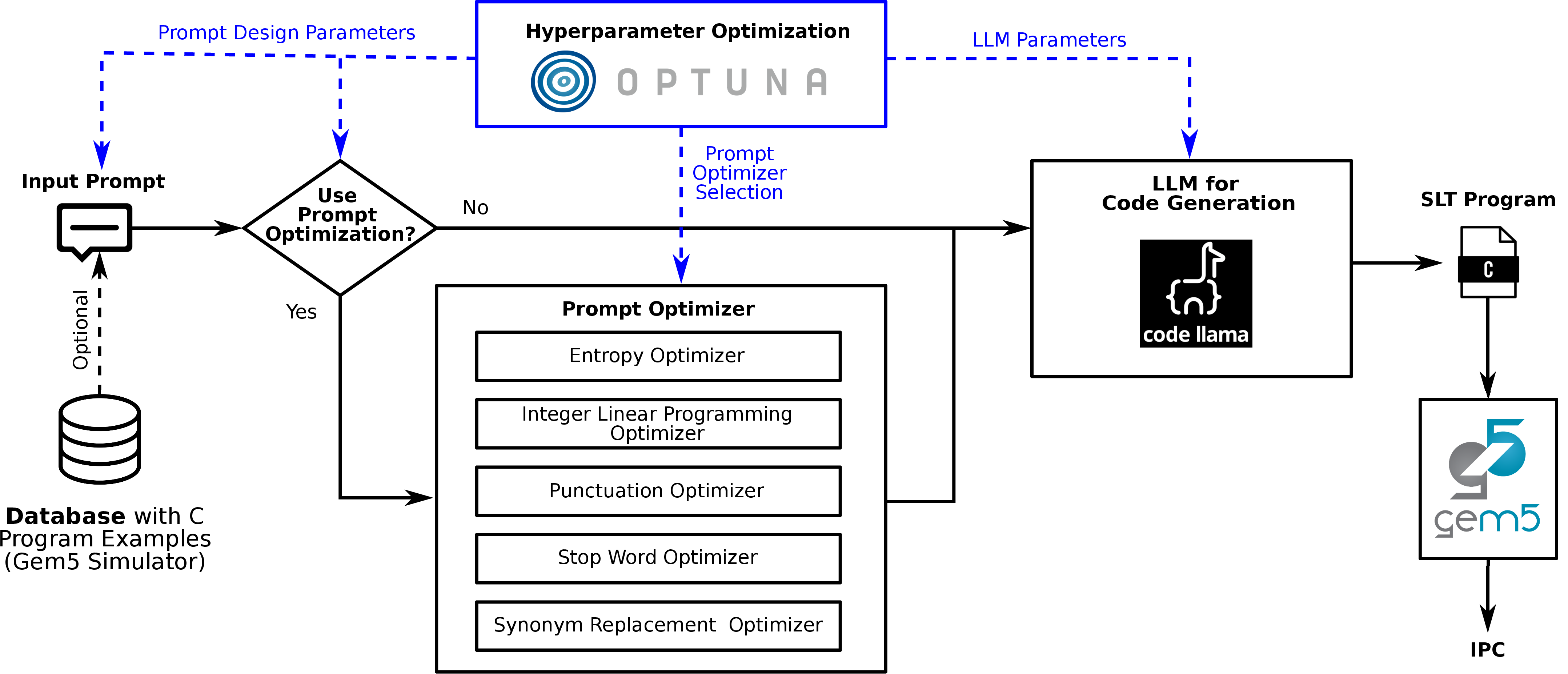}
    \caption{
        Overview of the proposed approach.
        The prompt is passed to Code Llama after optional optimization.
        We extract the generated code from its output.
        The code is executed in a Gem5 simulation, which will return the IPC.
    }
    \label{fig:overview_llm_slt}
\end{figure*}

\section{Proposed Approach}\label{sec:approach}

\Cref{fig:overview_llm_slt} shows the experimental setup.
We use the following input prompt:
\enquote{\texttt{<s>[INST]<<SYS>>}\textit{You are a C code generator.
Only respond with generated code and no explanation.
Do not justify the code.
Do not return C+incorrectays embed the generated code in Markdown code tags.
If a question does not make any sense, r it not factually coherent, explain why instead of answering something not correct.
If you don't know the answer to a question, please don't share false information.}
\texttt{<</SYS>>}
\textit{Write a single program that aims for a high number of instructions per cycle.
Don't forget to include all the necessary header files.
If you use math functions include math.h.}
\texttt{[/INST]}.}
The text in between \texttt{<<SYS>>} and \texttt{<</SYS>>} is a so-called system prompt.
It sets the context for Code Llama and thus does not need to be repeated for every consecutive user prompt.
The other text is called the user prompt and is the actual instruction for Code Llama.
The \texttt{[INST][/INST]} pair indicates the complete prompt, as future prompts can contain previous answers outside those to keep a conversation history.

We specifically ask Code Llama to return the code in Markdown code tags (surrounded by triple backticks) to allow us to tolerate more outputs.
Code Llama seems to sometimes ignore the instructions to not provide any explanations on the generated code.
Extracting the code from these tags allows us to ignore these explanations.

We provide two hand-written code snippets that can be used to enhance the prompt.
They are presented to Code Llama as example programs that compile and run.
However, we did not add their IPC to the prompt.
The assumption is that Code Llama will use these examples as a reference and derive a new program from these, but we did not ask it explicitly to do so.

We use \texttt{prompt-optimizer}~\cite{kumar_vaibkumrprompt-optimizer_2024} for prompt optimization.
It can be set up to use multiple algorithms, for example, based on entropy (infill low surprising words in the context) or replacing words with synonyms.
\Cref{fig:overview_llm_slt} shows all available algorithms.
We optimize the user and the system prompt separately to use the LLM most efficiently.

Finally, we use \texttt{Optuna}~\cite{Akiba2019Optuna} for hyperparameter optimization to maximize IPC.
It implements the Bayesian optimization method, a Parzen-Tree Estimation-based algorithm.
We let it optimize the following parameters: Temperature and Repetition Penalty of Code Llama.
We also allow it to enable or disable prompt optimization, let it choose one or a combination of two algorithms, and decide if the system prompt should be optimized additionally or only the user prompt.
Finally, we let it decide if the two hand-written code examples should be included in the input prompt.
The basic idea is to automatically find a constellation of parameters that consistently provide the best results based on IPC as feedback.

\section{Experimental Results}\label{sec:results}

In our experiment, we are deploying the \texttt{Instruct} Code Llama model with 13 billion parameters.
This model is optimized for generating code according to instructions given in the input prompt.
To evaluate the generated snippets, we use \texttt{Gem5} to simulate a 3-issue wide, super-scalar, out-of-order processor.
We let the simulation run for one billion ticks, which corresponds to a simulated millisecond.
\texttt{Optuna} runs 1000 trials to determine the optimal parameters.
We execute five runs to determine the stability of the hyperparameter optimization.

We are able to generate code snippets with a high IPC, 0.799607, which turns out to be as high as found by genetic programming using the same simulation model.
Throughout the runs, the best parameters were inconsistent regarding temperature and repetition penalty.
However, in general, there is a trend to not use prompt optimization.
Additionally, it turns out that including the pre-existing code examples in the prompt has been deemed non-beneficial as well.

1,002 snippets of the 5,000 generated snippets did not compile or crashed the \texttt{Gem5} simulation.
There are multiple reasons for the compilation to fail, for example, Code Llama responding that it doesn't understand the query, generating C++ or x86 Assembly instead or an incorrect output format that we cannot parse correctly.
In general, it appears that Code Llama does not understand the concept of IPC and how it applies to generated code.
Moreover, it assumes that it is generating code for an x86 machine.

Other cases of failing snippets are caused by improperly placed code tags; they either have been doubled up or the closing tag is missing entirely.
We do accept outputs that only contain code snippets without the code tags as well.
The latter also points to incomplete snippets, as Code Llama seems to have reached the maximum number of new tokens the model is able to generate.
The reason for the crashes is due to illegal memory accesses caused by, e.g., null pointer or out-of-bounds accesses.

We calculate the pass@k metric for \(k = {1,5}\) to evaluate the performance of the SLT program generation~\cite{chen2021evaluating}.
It describes the probability that at least one of the top k generated code snippets is a pass.
In this work, a pass is defined as a snippet that compiles and does not crash the simulation.
For \(k = 1\) we have \(79.96\%\) and for \(k = 5\) it is \(99.97\%\).
From these values, we can infer that the chance to get a snippet that is compiling and not crashing the simulation is four in five.

The above definition of pass does not include the IPC, which would most likely reduce the probabilities.
Let pass be a snippet that reaches a snippet with IPC at least \(0.5\).
Then for \(k = 1\) we would reach a chance of \(70.74\%\) and for \(k=5\) we have \(99.79\%\).
The probability of generating a snippet with an IPC higher than \(0.5\) is, in the case of pass@5, very close to the probability of generating a valid snippet in the first place, with a difference of only \(0.18\%\).
For pass@1, the difference of \(9.22\%\) is more significant, however, still high.

The prompt presented in~\Cref{sec:approach} has been developed over time from a simple prompt asking Code Llama to generate a program with high IPC since it produced a high number of failing programs.
This prompt showed the best results in our experiments.
However, we assume that more prompt engineering is required to further improve the program generation.

\section{Conclusion and Future Work}\label{sec:conclusion}
In this work, we conducted a feasibility study on the performance of pre-trained LLMs for SLT program generation.
Our experiments indicate that careful engineering of the input prompt is essential, especially if we do not fine-tune the LLM to generate task-specific SLT programs that target, for example, non-functional properties like IPC.
Future work will include fine-tuning the LLM and further prompt engineering, including instructing the LLM to explicitly derive new snippets from given examples to improve its capabilities and output.

\section*{Acknowledgements}
\addcontentsline{toc}{section}{Acknowledgements}
This work was supported by Advantest as part of the Graduate School ``Intelligent Methods for Test and Reliability'' (GS-IMTR) at the University of Stuttgart.

\printbibliography

\end{document}